\documentclass[aps,11pt,prc,preprint,superscriptaddress,nofootinbib]{revtex4}

\usepackage[usenames]{color}
\usepackage{graphicx}
\usepackage{amsmath}
\usepackage{amsfonts}
\usepackage{amssymb}
\usepackage{mathrsfs}
\usepackage{bm}
\usepackage{verbatim}

\newcommand{\beq}{\begin{equation}}
\newcommand{\eeq}{\end{equation}}
\newcommand{\bea}{\begin{eqnarray}}
\newcommand{\eea}{\end{eqnarray}}

\newcommand{\grad}{\vec{\nabla}}

\newcommand{\mlo}{M_{\text{lo}}}
\newcommand{\mhi}{M_{\text{hi}}}
\newcommand{\chn}[3]{{{}^{#1}{#2}_{#3}}}
\newcommand{\cs}[2]{\chn{#1}{S}{#2}}
\newcommand{\cp}[2]{\chn{#1}{P}{#2}}
\newcommand{\cd}[2]{\chn{#1}{D}{#2}}
\newcommand{\cf}[2]{\chn{#1}{F}{#2}}
\newcommand{\csd}{{\cs{3}{1}-\cd{3}{1}}}
\newcommand{\cpf}{{\cp{3}{2}-\cf{3}{2}}}

\newcommand{\y}{\text{Y}}

\graphicspath{{../}}

\preprint{JLAB-THY-12-1495}
\preprint{INT-PUB-12-001}

\begin{document}

\title{Short-range nuclear forces in singlet channels}
\author{Bingwei Long}
\email{bingwei@jlab.org}
\affiliation{Excited Baryon Analysis Center (EBAC), Jefferson Laboratory, 12000 Jefferson
Avenue, Newport News, VA 23606, USA}
\author{C.-J. Yang}
\email{cjyang@email.arizona.edu}
\affiliation{Department of Physics, University of Arizona, Tucson, AZ 85721, USA}

\date{\today}

\begin{abstract}
Continuing our effort to build a consistent power counting for chiral nuclear effective field theory (EFT), we discuss the subleading contact interactions, or counterterms, in the singlet channels of nucleon-nucleon scattering, with renormalization group invariance as the constraint. We argue that the rather large cutoff error of the leading amplitude requires $\mathcal{O}(Q)$ of the EFT expansion to be nonvanishing, contrary to Weinberg's original power counting. This, together with the ultraviolet divergences of two-pion exchanges in the distorted-wave expansion, leads to enhancement of the $\cs{1}{0}$ counterterms and results in a pionless-theory-like power counting for the singlet channels.
\end{abstract}

\maketitle

\section{Introduction\label{sec_intro}}

Power counting is one of the essential ingredients of any effective field theory (EFT), which not only keeps track of an infinite number of operators and Feynman diagrams but also estimates \textit{a priori} the neglected contributions for a given order. Naive dimensional analysis (NDA), a cornerstone of Weinberg's original power counting~\cite{Weinberg:1990-1991} (WPC) for few-nucleon systems, is often employed to assess the size of coupling constants: each derivative on the Lagrangian terms is always suppressed by the underlying scale of chiral EFT, $\mhi$. Though phenomenologically successful~\cite{Ordonez:1993-1995, Epelbaum:1998ka-1999dj, Epelbaum:2004fk, Entem:2001cg, Entem:2002sf}, WPC has been shown to be inconsistent with the principle of renormalization group (RG) invariance~\cite{Kaplan:1996xu, Beane:2000wh, Beane:2001bc, Nogga:2005hy, Birse:2005um, saopaulo, YangOhio, Valderrama:2009ei, Valderrama:2011mv, Long:2011qx, Long:2011xw}, especially in the triplet channels where the singular attraction of one-pion exchange (OPE) calls for modifications to WPC at as early as leading order (LO)~\cite{Nogga:2005hy}. The issues of RG invariance are less acute in the singlet channels, since the LO amplitudes of WPC for these channels are indeed RG invariant~\cite{Kaplan:1996xu, Phillips:1996ae, PavonValderrama:2004nb, Nogga:2005hy}, if we ignore the complication of chiral extrapolation which will be dealt with in future publications. But a modification to WPC for the subleading counterterms in the singlet channels has been argued in Refs.~\cite{Barford:2002je, Birse:2005um, Valderrama:2009ei} to be necessary. Following our investigation of the triplet channels~\cite{Long:2011qx, Long:2011xw}, we use RG invariance as the guideline to study the subleading counterterms of the singlet $S$ and $P$ waves.

The EFT expansion of the $T$-matrix at low energies has the quintessential form
\begin{equation}
T = \sum_{n} \left(\frac{Q}{\mhi}\right)^n \mathcal{F}_n\left(\frac{Q}{\mlo}\right) \, ,
\end{equation} 
where $Q$ denotes generically external momenta, $n$ the counting index, $\mlo$ low-energy mass scales, and $F_n(x)$ the nonanalytic functions from loop integrals. Even though the nonperturbative unitarity requires any nonrelativistic, nonperturbative $T$-matrix to scale as $Q^{-1}$, we \emph{choose} to label LO as $\mathcal{O}(1)$ so that one does not need to change the standard chiral power counting for irreducible pion exchange diagrams; e.g., OPE is $\mathcal{O}(1)$. Subleading orders of the EFT expansion are labeled by their relative correction to LO, i.e., next-to-leading order (NLO) by $\mathcal{O}(Q/M_{\text{hi}})$ or $\mathcal{O}(Q)$ for short, and next-to-next-to-leading order (NNLO) by $\mathcal{O}(Q^{2}/M_{\text{hi}}^{2})$ or $\mathcal{O}(Q^2)$, and so on.

In any EFT calculations employing a ultraviolet (UV) momentum cutoff $\Lambda$, the cutoff independence--- RG invariance--- of the $T$-matrix is usually imperfect at a given order. There usually exists a residual cutoff dependence, or a cutoff error, of the $T$-matrix with a large but finite cutoff, which vanishes though as $\Lambda \to \infty$. Seen in this light, RG invariance needs a little more careful interpretation. Because (i) the cutoff error is \emph{part} of the theoretical uncertainty at a given order and (ii) the theoretical uncertainty is, by definition, of the same order as the next-order EFT correction, the cutoff error must be smaller than or of the same order as the next-order correction. This seemingly trivial statement constrains power counting in a nontrivial way when certain order is considered vanishing. Of our interest is WPC, in which the $\mathcal{O}(Q)$ corrections have long been deemed to be zero; thus, the theoretical uncertainty for LO is considered by WPC to be $\mathcal{O}(Q^2)$. It follows that the LO cutoff error should vanish at least as fast as $Q^2/\Lambda^2$:
\begin{equation}
T^{(0)}(Q; \Lambda) - T^{(0)}(Q; \infty) \lesssim \left(\frac{Q}{\Lambda}\right)^{2}
\, .
\label{eqn_WPCLOres}
\end{equation}
While this is the case for the triplet channels~\cite{Beane:2000wh, Beane:2001bc, PavonValderrama:2007nu}, we will show that it is not for $\cs{1}{0}$, which forces us to modify WPC for $\cs{1}{0}$ at subleading orders even though its LO satisfies RG invariance.

If a counterterm is not required by RG invariance at $\mathcal{O}(Q^n)$ but is counted $\mathcal{O}(Q^n)$ in NDA, we will follow NDA to power count that counterterm. In other words, we do not minimize the number of counterterms at a given order using RG invariance as the criterion~\cite{bira-private, Long:2011xw}. The rationale for this is as follows. RG analysis does not study the degrees of freedom that are not built in the effective Lagrangian, such as non-Goldstone bosons and/or very heavy excited baryon states. To avoid underestimating the contributions of these degrees of freedom to counterterms, we set the minimal size of counterterms as the one given by NDA as long as doing so does not violate RG invariance.

We will treat subleading potentials as perturbations on top of the LO $T$-matrix, which is nonperturbative iteration of the LO potential. This is sometimes casually called perturbative renormalization. Underlying this approach is the point of view that power counting should be done at the level of physical observables; in our case, the on-shell scattering amplitude. Infrared enhancement due to nucleon intermediate states should be incorporated into the power counting~\cite{Kaplan:1998tg, vanKolck:1999mw} rather than be used as the pretext to settle for WPC. However, this is not to say that a consistent power counting for the nonperturbative approach, in which the full iteration of the whole potential is performed, cannot be found. For developments in this direction, we refer the reader to Refs.~\cite{Epelbaum:2006pt, Epelbaum:2009sd, Epelbaum:2012vx}.

Our paper is structured as follows. In Sec.~\ref{sec_SLC}, the subleading counterterms are classified into three categories according to the diagrams that drive their RG evolution. After a short review of LO in the singlet channels, we discuss in Sec.~\ref{sec_LOandNLO} the rather large cutoff error of the LO $\cs{1}{0}$ amplitude and the consequence of that for power counting. We analyze in Sec.~\ref{sec_NNLO} how the nonvanishing $\mathcal{O}(Q)$ counterterm and two-pion exchanges contribute to $\mathcal{O}(Q^2)$. This is followed by a discussion and a conclusion offered in Sec.~\ref{sec_discussion}.

\section{Evolution of subleading counterterms\label{sec_SLC}}

The LO $T$-matrix, $T^{(0)}$, arises from the full iteration of OPE (in low partial waves) and a set of counterterms that ensure the RG invariance, which means that $T^{(0)}(k; \Lambda)$ is independent of $\Lambda$ when $\Lambda \gg k$, where $k$ is the magnitude of the center-of-mass momentum. Subleading orders are given by perturbative insertions of higher-derivative counterterms and/or irreducible multiple-pion exchanges into LO. Although renormalization at LO~\cite{Beane:2000fi, PavonValderrama:2004nb} is far more intricate, we expect to have a better visualization of renormalization at subleading orders by forming a fairly simple correspondence between a loop diagram and the counterterm to subtract its UV divergences, much like ones that exist in perturbative EFTs. In Wilson's language of RG analysis, this counterterm is the one that ``evolves'' most significantly when the cutoff of the loop diagram is rescaled from $\Lambda$ to a smaller value, $\Lambda'$, but remains large in the sense $\Lambda' \gg k$. It will help our discussion to classify loop diagrams and their corresponding counterterms into the following three categories.

\subsection{Residual counterterms\label{sec_SLC_res}}

Diagrams in the first class are actually those of $T^{(0)}$. Of course the LO contact operators, by definition, are the counterterms to renormalize $T^{(0)}$; however, in order to systematically remove the residual cutoff dependence of $T^{(0)}$, one must take account of the contact operators with more derivatives than the leading one. We call those higher-derivative operators the residual counterterms for $T^{(0)}$. Even though they are not as important as the LO counterterms or OPE, they might be more important than multiple-pion exchanges which start to contribute at $\mathcal{O}(Q^2)$. It is important for us to find a way to estimate their sizes before calculations are carried out.

The authors of Refs.~\cite{Barford:2002je, Birse:2005um, Birse:2009my} have attempted to analyze both LO and residual counterterms using the Wilson RG equation, with OPE as the only long-range force. This is a very difficult task and several assumptions were made in Refs.~\cite{Barford:2002je, Birse:2005um, Birse:2009my}. Energy and momentum dependences of contact operators were assumed to be independent of each other. But we know that, when treated as perturbations on top of LO, they can be related by the equation of motion. The RG invariance of the off-shell $T$-matrix was imposed, though only the on-shell quantities need to be RG invariant. While this excessive requirement cannot be deemed wrong, one may be concerned that the resulting power counting demands more counterterms than necessary, only to ensure the RG invariance of the off-shell part of the $T$-matrix. \footnote{
Reference~\cite{Entem:2009mf} showed that the half-off-shell partial-wave $T$-matrix for singular potentials, $T(p', k; k)$, is well-defined as $\Lambda \to \infty$. But the presumption of RG equations in Refs.~\cite{Barford:2002je, Birse:2005um} seems to be stronger: for any $p'$ between $k$ and (finite) $\Lambda$, the cutoff dependence of $T(p', k; k)$ needs to vanish uniformly at a rate independent of $p'$ so that when $\Lambda$ varies the variation of the integral in the Lippmann-Schwinger equation is dominated by the contribution due to the varied endpoint, as opposed to the contribution due to the functional change of $T(p', k; k)$ (as a function of $p'$).} What is most debatable is perhaps the existence of an infrared fixed-point solution to the RG equation, around which the power counting is obtained. Although this appears to be reasonable in the singlet channels, it is clearly at odds with the running of counterterms in the attractive triplet channels, where a limit-cycle-like behavior was observed~\cite{Beane:2000wh, Nogga:2005hy}.

Instead of performing a comprehensive analysis of the RG equation, we \emph{impose} a narrower definition for residual counterterms: the correction brought by the residual counterterm is of the same size as the cutoff error $T^{(0)}(k; \Lambda) - T^{(0)}(k; \infty)$. For example, the LO cutoff error in $\csd$ is found to be $\mathcal{O}(k^2\mlo^{1/2}\Lambda^{-5/2})$~\cite{Beane:2000wh, Beane:2001bc, PavonValderrama:2007nu}. Thus, the correction due to the $\cs{3}{1}$ residual counterterm is rated as $\mathcal{O}(Q^2\mlo^{1/2}\Lambda^{-5/2})$, less than the leading two-pion exchange (TPE0), which is $\mathcal{O}(Q^2/\mhi^2)$.

It should now become apparent that we use the residual counterterm merely as a mnemonic device to reflect the order of magnitude of the LO cutoff error. With this notion we can interpret the inequality~\eqref{eqn_WPCLOres} as follows: WPC requires the residual counterterm to be no more important than TPE0. It is indeed true for the triplet channels, but, as we will show, it is not so for $\cs{1}{0}$.

\subsection{Primordial counterterms}

Better known is the second class of diagrams: irreducible multiple-pion exchanges evaluated in the plane, or free spherical, wave basis. Primordial counterterms~\cite{Long:2011xw} are the contact operators necessary to subtract the divergences of these pion-exchange diagrams.

On the basis of ``naturalness,'' the power counting of primordial counterterms should be the same as that of the pion-exchange diagram in question, which is reliably handled by WPC. For example, TPE0 is $\mathcal{O}(Q^2)$ and its primordial counterterm, a second-order polynomial in momenta, is counted $\mathcal{O}(Q^2)$ as well.

\subsection{Distorted-wave counterterms}

Diagrams of the third class are insertions of irreducible multiple-pion exchanges into LO; that is, pion exchanges sandwiched between distorted waves--- the LO wave functions, $\psi_k$. The counterterms to absorb the divergences of these diagrams are called by us distorted-wave counterterms.

Coordinate space provides the best stage for qualitative discussion of distorted-wave UV divergences. Any reasonable UV regulator will roughly separate radial coordinates into two parts: the inside, $0 < r \lesssim \Lambda^{-1}$, and the outside, $\Lambda^{-1} \lesssim r < \infty$. Details of the regulator decide how sharp the separation is. When contact interactions are present at LO, such as in both $S$ waves, the inside and outside parts of the LO wave function, $\psi_k^\text{in}$ and $\psi_k^\text{out}$, are respectively subject to the LO contact potential and OPE. The LO contact potential and OPE have different short-distance structures, which result in different short-distance behaviors of $\psi_k^\text{in}$ and $\psi_k^\text{out}$. For instance, as shown in, e.g., Refs.~\cite{Phillips:1996ae, Scaldeferri:1996nx}, $\psi_k^\text{out}(r)$ in $\cs{1}{0}$ has an irregular component diverging like $\sim 1/r$ near $r \sim \Lambda^{-1}$, which differs very much from a free spherical wave. On the other hand, since the LO contact potential is always well defined upon regularization, the inside wave function $\psi_k^\text{in}$ is not drastically different from a free spherical wave.

The distorted-wave matrix element of a subleading contact potential, $\langle \psi_k| V_S^{\text{sub}} | \psi_k \rangle$, is dominated by the integration over the inside region. Since $\psi_k^\text{in}(r)$ behaves similarly to a free wave at short distance, $\langle \psi_k| V_S^{\text{sub}} | \psi_k \rangle$ is expected to be as UV singular as its free-wave counterpart, i.e., $\langle \text{plane wave}| V_S^{\text{sub}} | \text{plane wave} \rangle$.

On the other hand, the distorted-wave matrix element of a subleading long-range potential, $\langle \psi_k| V_L^{\text{sub}} | \psi_k \rangle$, is mostly decided by the integration of $\psi_k^\text{out}(r)$. The irregular component of $\psi_k^\text{out}(r)$, if present, could make $\langle \psi_k| V_L^{\text{sub}} | \psi_k \rangle$ more divergent than its free-wave counterpart. Therefore, the primordial counterterm that renormalizes a given multiple-pion exchange in the plane-wave basis may no longer renormalize the same pion exchange between distorted waves. If it does not, the distorted-wave counterterm will have to be more singular than its primordial counterpart; that is, it will have more derivatives. This indeed happens to a toy model considered in Ref.~\cite{Long:2007vp} and in $\cp{3}{0}$, $\cpf$~\cite{Valderrama:2009ei, Valderrama:2011mv, Long:2011qx}, and $\cs{1}{0}$~\cite{Valderrama:2009ei}, among possibly other channels of $NN$ scattering. We will reproduce in Sec.~\ref{sec_NNLO} the distorted-wave enhancement of the $\cs{1}{0}$ subleading counterterms, which was first shown in Ref.~\cite{Valderrama:2009ei}.

The distorted-wave multiple-pion exchanges, $\langle \psi_k| V_L^{\text{sub}} | \psi_k \rangle$, are power counted the same as the free-wave matrix elements because it has been established at LO that any number of insertions of $V^{(0)}$--- the LO potential--- does not enhance or diminish the amplitude. Based on, again, naturalness, it follows that the distorted-wave counterterm is power counted the same as its primordial counterpart, even though the distorted-wave counterterm may have more derivatives.

\section{LO and its residual counterterms\label{sec_LOandNLO}}

We consider first $^1S_0$. The LO amplitude is constructed by resumming $V^{(0)}$, which is a constant counterterm $C_\cs{1}{0}$ plus OPE,
\begin{equation}
V^{(0)}(q) = V_\pi(q) + C_{\cs{1}{0}} \, ,
\end{equation}
where
\begin{equation}
V_\pi(q) = \frac{g_A^2}{4f_\pi^2} \frac{q^2}{q^2 + m_\pi^2}\, ,
\label{eqn_Vpiq}
\end{equation}
with $\vec{p}\,'$ ($\vec{p}\,$) being the outgoing (incoming) momentum in the center-of-mass frame, $\vec{q} \equiv \vec{p}\,' - \vec{p}$, $g_A = 1.26$, and $f_\pi = 92.4$ MeV. We can redefine $C_\cs{1}{0}$ such that the pointlike piece embedded in OPE is separated from the Yukawa potential, and rewrite long- and short-range potentials for $\cs{1}{0}$, respectively, as
\begin{equation}
V_Y(q) = -\frac{4\pi}{m_N} \frac{\alpha_\pi m_\pi^2}{q^2 + m_\pi^2}\, , \quad V_S^{(0)} = C^{(0)} \, ,
\end{equation} 
where $\alpha_\pi \equiv g_A^2 m_N/16\pi f_\pi^2 \sim (290 \text{MeV})^{-1}$ and $4\pi/m_N$ is a common factor of nonrelativistic Feynman amplitudes. Here we have dropped the subscript $\cs{1}{0}$ to simplify the notation. $C$ has been formally expanded in anticipation that the running of $C(\Lambda)$ with respect to $\Lambda$ could be modified at each order,
\begin{equation}
C(\Lambda) = C^{(0)}(\Lambda) + C^{(1)}(\Lambda) + \cdots \, ,
\end{equation}
though the number of physical inputs to determine $C$ remains one (or stated differently, the boundary condition for the RG flow of $C(\Lambda)$ remains fixed). Barring fine-tuning, the power counting of 
(renormalized) $C$~\cite{vanKolck:1999mw, Bedaque:2002mn, Barford:2002je} is decided by the pointlike piece of OPE:
\begin{equation}
C_R \sim \frac{4\pi}{m_N} \frac{1}{\mlo} \, ,
\label{eqn_CR}
\end{equation} 
where $Q \sim \mlo \sim \alpha_{\pi}^{-1}$. The relatively large size of $\alpha_\pi^{-1}$, compared with $m_\pi$, is crucial for the singlet-channel success of the scheme by Kaplan, Savage, and Weiss (KSW)~\cite{Kaplan:1998tg, Fleming:1999ee}, in which OPE is treated perturbatively. But here we take the view point that $\alpha_\pi^{-1}$ is an infrared mass scale because (i) it is still smaller than $\mhi$ and (ii) the perturbative Yukawa works less well when $m_\pi$ takes a larger value but still stays within the validity of chiral EFT.

\begin{figure}
\centering
\includegraphics[scale=0.4]{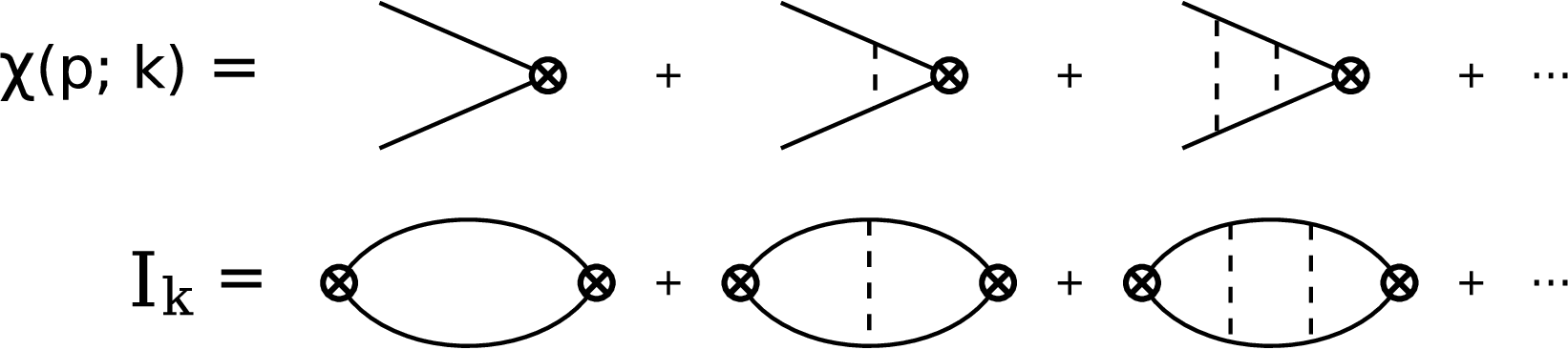}
\caption{Diagrammatic representation of $\chi(p; k)$ and $I_k$. Here the solid (dashed) lines represent the 
nucleon (pion) propagator, and the crossed circles represent no interaction.
\label{fig_I_k}}
\centering
\end{figure}

We will use conventional numerical methods to carry out actual calculations (see Sec.~\ref{sec_NNLO_num}), but we can study the UV divergences analytically using the elegant machinery developed in Ref.~\cite{Kaplan:1996xu}. First, we define the resummed Yukawa amplitude: 
\begin{equation}
T_\y(\vec{p}\,', \vec{p}\,; k) = V_\y(|\vec{p}\,' - \vec{p}\,|) + \int \frac{d^3l}{(2\pi)^3} V_\y(|\vec{p}\,' - \vec{l}\,|)\, \frac{T_\y(\vec{l}, \vec{p}\,; k)}{E - \frac{l^2}{m_N} + i\epsilon} \, ,
\end{equation} 
where $E \equiv k^2/m_N$ is the center-of-mass energy. While the LO $P$ and higher wave amplitudes are given solely by the resummed Yukawa amplitude, the LO $S$-wave amplitude requires summing up insertions of $C^{(0)}$ to all orders, which is eventually given by~\cite{Kaplan:1996xu}
\begin{equation}
T^{(0)}(\vec{p}\,', \vec{p}\,; k) = T_\y(\vec{p}\,', \vec{p}\,; k) + \frac{\chi(p'; k)\chi(p; k)}{{(C^{(0)})}^{-1} - I_k} \, ,
\label{eqn_TLO}
\end{equation} 
where
\begin{align}
\chi(p; k) &= 1 + \int \frac{d^3l}{(2\pi)^3} \frac{T_\y(\vec{l}, \vec{p}\,; k)}{E - \frac{{l}^2}{m_N} + i\epsilon} \label{eqn_chiE} \, , \\
I_k &= \int \frac{d^3l}{(2\pi)^3} \frac{\chi(l; k)}{E - \frac{{l}^2}{m_N} + i\epsilon} \label{eqn_IE}  \, .
\end{align} 
Figure~\ref{fig_I_k} shows the diagrams that, when resummed, represent $I_k$ and $\chi(p; k)$. The power counting of $I_k$ and $\chi(k; k)$ will follow, e.g., that of the first diagram of their resummation series: $I_k \sim \frac{m_N Q}{4\pi}$ and $\chi(k; k) \sim 1$. Equation~\eqref{eqn_TLO} is exactly correct only when $V^{(0)}$ is dimensionally regularized~\cite{Kaplan:1996xu} or regularized by a separable cutoff regulator:
\begin{equation}
V^{(0)}_\Lambda(\vec{p}\,', \vec{p}\,) \equiv f_R\left(\frac{{p'}^2}{\Lambda^2}\right)  V^{(0)}\left(|\vec{p}\,' - \vec{p}\,| \right) f_R\left(\frac{{p}^2}{\Lambda^2}\right) \, .
\end{equation}
For a more general regulator, Eq.~\eqref{eqn_TLO} is true only at $\Lambda \to \infty$. We will assume in studying the LO cutoff error that a separable regulator is used.

Although we cannot calculate analytically $I_k$, $\chi(p; k)$, or $T_Y$, the dominant UV divergences can be captured by noticing that every insertion of $V_\y$--- combined with the Schr\"odinger propagator--- suppresses
the UV divergences by $1/\Lambda$~\cite{Kaplan:1996xu}. Therefore, the most significant cutoff dependences of $I_k$ are in the first two diagrams of the lower row in Fig.~\ref{fig_I_k}: the first is linear and the second is logarithmic in $\Lambda$,
\begin{equation}
\frac{4\pi}{m_N} I_k = \frac{4\pi}{m_N} \left(I_\Lambda + I_k^R\right)  +  \beta_2(k, \kappa_\pi, m_\pi^2)\, \frac{k^2}{\Lambda} + \mathcal{O}\left(\frac{\kappa_\pi k^2}{\Lambda^2}\right) \label{eqn_IEres} \, ,
\end{equation}
with
\begin{equation}
\frac{4\pi}{m_N} I_\Lambda \equiv \beta_0\, \Lambda + \beta_1 \kappa_\pi \ln\left(\frac{\Lambda}{\kappa_\pi}\right) \, . \label{eqn_ILambda}
\end{equation}
Here $I_k^R$ is the finite part, $\kappa_\pi = \alpha_\pi m_\pi^2$, and
$\beta_i$ are dimensionless and depend on the details of the UV regulator. While $\beta_0$ and $\beta_1$ are numbers coming out of the first two diagrams, the $1/\Lambda$ cutoff dependence, $\beta_2$, receives contributions from all the diagrams; therefore, $\beta_2$ is a nonperturbative function of $k$, $\kappa_\pi$, and $m_\pi^2$. In the spirit of keeping track of $1/\Lambda$ cutoff dependences, one finds that
\begin{equation}
\chi(k; k) = \chi_k^R + \gamma_2(k, \kappa_\pi, m_\pi^2) \frac{\kappa_\pi}{\Lambda} + \mathcal{O}\left(\frac{\kappa_\pi^2}{\Lambda^2}\right) \label{eqn_chiEres} \, ,
\end{equation}
where $\chi_k^R$ is the finite part and $\gamma_2$ is another dimensionless function. As for $T_Y$, we notice that its residual cutoff dependence is only $\mathcal{O}(\kappa_\pi k^2/\Lambda^3)$.

We would like to have available the coordinate-space form of the LO wave function, which will be useful in analyzing the distorted-wave counterterms for TPEs. Since they are somewhat out of the main line of our discussion, the relevant expressions concerning the wave function are relegated to Appendix~\ref{app_wf}.

Introducing the renormalized coupling $C_R$, such that
\begin{equation}
C_R^{-1} = {(C^{(0)})}^{-1} - I_\Lambda \, , \label{eqn_CRdef}
\end{equation}
we can rewrite the on-shell $T$-matrix as
\begin{equation}
T^{(0)}(\vec{k}, \vec{k}; k) = T_Y(\vec{k}, \vec{k}\,; k) + \frac{{\chi(k; k)}^2}{C_R^{-1} - I_k^R + \mathcal{O}\left(\frac{m_N k^2}{4\pi \Lambda}\right) } \, .
\label{eqn_TLOR}
\end{equation}
\footnote{If we had regularized the infrared end of the logarithmic divergence in $I_\Lambda$ by an arbitrary mass $\mu$ instead of $\kappa_\pi$, $C_R$ and $I_k^R$ would depend on $\mu$ in a way that $(C_R^{-1} - I_k^R)$ does not.}
As $\Lambda \to \infty$, Eq.~\eqref{eqn_TLOR} is no more than reproducing one of the results in Ref.~\cite{Kaplan:1996xu}. However, one can go further and infer from the cutoff error some information about the subleading $\cs{1}{0}$ counterterms. Using Eqs.~\eqref{eqn_IEres} and \eqref{eqn_chiEres}, power counting \eqref{eqn_CR}, $I_k \sim \frac{m_N Q}{4\pi}$, and $\chi(k; k) \sim 1$, one finds, as promised, that the cutoff error is $\mathcal{O}(k^2/\mlo \Lambda)$:
\begin{equation}
\frac{{\chi_k^R}^2}{C_R^{-1} - I_k^R}\left[2\frac{\gamma_2}{\chi_k^R} \frac{\kappa_\pi}{\Lambda} + \beta_2 \frac{k^2}{\frac{4\pi}{m_N} \left(C_R^{-1} - I_k^R\right) \Lambda} \right] \, .\label{eqn_LOcutoff}
\end{equation} 
Therefore, the theoretical uncertainty of $T^{(0)}$ must be $\mathcal{O}(Q)$, or equivalently, the residual counterterm--- the two-derivative $\cs{1}{0}$ contact operator $D/2({p'}^2 + p^2)$--- is $\mathcal{O}(Q)$, following our definition of residual counterterms in Sec.~\ref{sec_SLC_res}. On the other hand, since pion corrections do not start until $\mathcal{O}(Q^2)$, a nonvanishing $\mathcal{O}(Q)$ can only be one insertion of the $D$ term:
\begin{equation}
T^{(1)} = \left(1 + T^{(0)}G\right)V_S^{(1)}\left(G T^{(0)} + 1 \right) \, ,\label{eqn_T1abs}
\end{equation} 
where $V_S^{(1)}$ denotes the short-range part of the $\mathcal{O}(Q)$ potential,
\begin{equation}
\langle \cs{1}{0} | V_S^{(1)} | \cs{1}{0} \rangle = C^{(1)} + \frac{D^{(0)}}{2}({p'}^2 + p^2) \, , \label{eqn_VS1}
\end{equation} 
and $G$ is the Schr\"odinger propagator.

If the loop integrals in $\chi(k; k)$~\eqref{eqn_chiE} and $I_k$~\eqref{eqn_IE} are dimensionally regularized, the LO cutoff error~\eqref{eqn_LOcutoff} vanishes.  But it is model-dependent thinking to count on specific regulators to be superior and to ignore the cutoff errors that arise with other regulators.

Higher partial-wave amplitudes are decided by $T_Y$ alone. The quite small cutoff error of $T_Y$, $\mathcal{O}(\kappa_\pi k^2/\Lambda^3)$, means that the residual counterterm for $\cp{1}{1}$ is no more important than $\mathcal{O}(Q^2)$ and that WPC does not need to change for $\cp{1}{1}$:
\begin{equation}
\langle \cp{1}{1} | V_S^{(1)} | \cp{1}{1} \rangle = 0 \, .
\label{eqn_1p1T1}
\end{equation}

Starting from $\mathcal{O}(Q)$, we no longer enjoy the ease of keeping track of residual cutoff dependence as we did for the LO $\cs{1}{0}$ amplitude. The main technical reason is related to the fact that the separable cutoff regulator used at LO causes $V_\Lambda^{(0)}$ to be nonlocal in coordinate space at short distance, as explained in more detail in Appendices~\ref{app_wf} and \ref{app_tricks}. Fortunately, cutoff errors at subleading orders are not expected to constrain power counting in the way they did for LO, because after multiple-pion exchanges kick in at $\mathcal{O}(Q^2)$ we do not expect any subleading order to vanish. However, the difference between dimensional regularization and cutoff regulators with $\Lambda \to \infty$ can still be relevant for power counting, as will be seen in Sec.~\ref{sec_NNLO}.

In the limit $\Lambda \to \infty$, one can at least obtain formal expressions for insertions of subleading contact interactions, which are sufficient to see how divergences are subtracted. Through the steps shown in Appendix~\ref{app_tricks}, one can rewrite the $\mathcal{O}(Q)$ $\cs{1}{0}$ amplitude~\eqref{eqn_T1abs} in a more comprehensible form:
\begin{equation}
T^{(1)} = \frac{1}{\left(C^{(0)}\right)^2} \frac{{\chi_k^R}^2}{\left(C_R^{-1} - I_k^R\right)^2} \left\{ \left[C^{(1)} - D^{(0)} m_N \widetilde{V}^{(0)}(0) \right] + D^{(0)} k^2 \right\}  \, ,\label{eqn_T1B}
\end{equation} 
where
$\widetilde{V}^{(0)}(0)$ \eqref{eqn_VLOtilde} is the formal value of $V^{(0)}$ at the spacial origin. Since $C^{(1)}$ does not incorporate any new physical information, it is at our disposal to choose the value of $C^{(1)}$ as long as it helps renormalization. With
\begin{equation}
D_R \equiv \frac{D^{(0)}}{{C^{(0)}}^2} C_R^2 \, ,
\end{equation}
we choose the value of $C^{(1)}$ such that
\begin{equation}
\frac{C^{(1)}}{(C^{(0)})^2} C_R^2 - m_N D_R \widetilde{V}^{(0)}(0) = 0 \, .
\end{equation}
Now we can express $T^{(1)}$ in terms of renormalized quantities:
\begin{equation}
T^{(1)} = \frac{D_R}{C_R^2} \frac{k^2 {\chi_k^R}^2}{\left(C_R^{-1} - I_k^R\right)^2}
\, . \label{eqn_T1R}
\end{equation}
It is obvious that, for $T^{(1)}$ to be $\mathcal{O}(Q)$, the scaling of $D_R$ must be
\begin{equation}
D_R \sim \frac{4\pi}{m_N} \frac{1}{\mlo^2 \mhi} \, , \label{eqn_DRpc}
\end{equation}
in comparison with $C_R$~\eqref{eqn_CR}. The interpretation of $T^{(1)}$ becomes particularly simple in the chiral limit where $T_Y$ vanishes: the $D$ term plays the role of the effective range.

\section{$\mathcal{O}(Q^2)$\label{sec_NNLO}}

\subsection{$S$ Wave}

At $\mathcal{O}(Q^2)$ there are \emph{two} insertions of $V_S^{(1)}$ and one insertion of each of TPE0 (denoted by $V^{(0)}_{2\pi}$) and $V_S^{(2)}$, where $V_S^{(2)}$, before any higher-derivative counterterm is considered, includes at least the $\mathcal{O}(Q^2)$ corrections to $C(\Lambda)$ and $D(\Lambda)$: $C^{(2)}(\Lambda)$ and $D^{(1)}(\Lambda)$.

Two insertions of $V_S^{(1)}$ include integrals involving the LO interacting Green function: $\mathscr{G}_k \equiv G(1 + T^{(0)}G)$. The contributions of $C^{(2)}$, $D^{(1)}$, and two $V_S^{(1)}$'s are eventually summed up as
\begin{equation}
\begin{split}
T^{(2)}_{S} = \frac{{\chi_k^R}^2}{\left(C_R^{-1} - I_k^R\right)^2} \left[ \frac{D_R^2}{C_R^4} \frac{k^4}{\left(C_R^{-1} - I_k^R\right)} + \frac{D_R^2}{C_R^4} C^{(0)} k^4 + \left(\mathcal{A} + \mathcal{B} k^2 \right) \right] \, , \\
\end{split}
\label{eqn_T2S}
\end{equation} 
where
\begin{align}
\mathcal{A} &= - m_N \frac{D_R^2}{C_R^4} \left(C^{(0)}\right)^2 \left[m_N\widetilde{V}^{(0)}(0) \delta^{(3)}(0) - {\delta^{(3)}}^{\prime\prime}(0) \right] + \frac{C^{(2)}}{{C^{(0)}}^2} \, , \label{eqn_Adef} \\
\mathcal{B} &= - \frac{3}{4}m_N \frac{D_R^2}{C_R^4} \left(C^{(0)} \right)^2 \delta^{(3)}(0) + \frac{D^{(1)}}{{C^{(0)}}^2} \, . \label{eqn_Bdef}
\end{align} 
For the related computational details of the above equation (and Eqs.~\eqref{eqn_T2E} and \eqref{eqn_T2epsilon}), we refer the reader to Appendix~\ref{app_tricks}. The first term in the brackets of Eq.~\eqref{eqn_T2S} does not bring more information than $T^{(1)}$ \eqref{eqn_T1R}; it merely restores the unitarity up to $\mathcal{O}(Q^2)$. The second term has new structure, which becomes more apparent in the chiral limit where, since $V_Y$ vanishes, it resembles the shape parameter of a contact-only theory. We will casually refer to it below as the shape parameter term even away from the chiral limit. With power countings \eqref{eqn_CR} and \eqref{eqn_DRpc} and Eqs.~\eqref{eqn_ILambda} and \eqref{eqn_CRdef}, one sees that the shape parameter term vanishes for cutoff regulators in the large-$\Lambda$ limit where $C^{(0)} \propto 1/(\beta_0 \Lambda)$. With $\mathcal{A}$ and $\mathcal{B}$ made finite by $C^{(2)}$ and $D^{(1)}$, $T^{(2)}_S$ is well defined as $\Lambda \to \infty$.

However, if dimensional regularization with minimal subtraction was used to regularize $I_\Lambda$ \eqref{eqn_ILambda} (the first two diagrams of $I_k$ in Fig.~\ref{fig_I_k}), we are led to a different opinion on counting the shape parameter term. In the chiral limit, the Yukawa amplitude vanishes and we will have $C^{(0)} = C_R \propto 1/\mlo$, which means that the shape parameter term is an $\mathcal{O}(Q^2)$ contribution, in contrast to what happens with cutoff regulators. Since a consistent power counting does not discriminate against certain regularization schemes, we must add a (residual) four-derivative counterterm $E^{(0)}{p'}^2 p^2$ at $\mathcal{O}(Q^2)$ to absorb the regularization-scheme dependence. For finite $m_\pi$, the pole term stemming from dimensional regularization of the second diagram of $I_k$ in Fig.~\ref{fig_I_k} will cause $C^{(0)} = 1/({C_R}^{-1} - I_\Lambda)$ to vanish. In such a case the disparity between both regularization schemes is no longer a concern in regard to renormalization. Nevertheless, in order to have an easier transition to the chiral limit, we will count the residual $E$ as $\mathcal{O}(Q^2)$ for finite $m_\pi$ even though it is not strongly required by renormalizability. (Interestingly, as we will see soon, this decision is not crucial for power counting after all: $E$ will at any rate be required at $\mathcal{O}(Q^2)$ as the distorted-wave counterterm for two-pion exchanges, regardless of the value of $m_\pi$.) One insertion of the $E$ term yields
\begin{equation}
T^{(2)}_{E} = \frac{E^{(0)}}{{C^{(0)}}^2} \frac{{\chi_k^R}^2}{\left(C_R^{-1} - I_k^R\right)^2} \left[k^2 - m_N\widetilde{V}^{(0)}(0) \right]^2 \, . \label{eqn_T2E}
\end{equation}

Before considering TPE0, we note that the other four-derivative term $\mathscr{E}({p'}^4 + p^4)$, when treated as perturbation, is redundant:
\begin{equation}
T^{(2)}_{\mathscr{E}}
= \frac{\mathscr{E}}{{C^{(0)}}^2} \frac{{\chi_k^R}^2}{\left(C_R^{-1} - I_k^R\right)^2} \left\{ \left[k^2 - m_N\widetilde{V}^{(0)}(0)\right]^2 - m_N 
\grad^2{\widetilde{V}^{(0)}}(0)
\right\}\, .\label{eqn_T2epsilon}
\end{equation} 
With Eqs.~\eqref{eqn_T1R} and \eqref{eqn_T2E}, $T^{(2)}_\mathscr{E}$ can be expressed in the large $\Lambda$ limit as a combination of the $C$ and $E$ operators. A more general argument is of course the field redefinition inspired by the nucleon equation of motion~\cite{Beane:2000fi}. For a general cutoff regulator with finite $\Lambda$, ${p'}^4 + p^4$ is not necessarily equivalent to ${p'}^2p^2$. But their difference for finite $\Lambda$, in a consistent power counting, is no more significant than the cutoff error when either operator, but not both, is used.

The analytic part of $V_{2\pi}^{(0)}(q)$ is a second-order momentum polynomial, i.e., its primordial counterterm, which, when projected onto $\cs{1}{0}$, is nothing but the $D$ term. The insertion of the $D$ term into $T^{(0)}$ is shown in Sec.~\ref{sec_LOandNLO}. We now consider the matrix element of the nonanalytic part, which diverges as $r \to 0$ in coordinate space: $\widetilde{V}_{2\pi}^{(0)}(r) \sim 1/(\mhi^2 r^5)$. This was first shown in Ref.~\cite{Valderrama:2009ei}, though in a slightly different notation than ours.

This is perhaps most readily done in coordinate space where $V_{2\pi}^{(0)}$ is diagonal,
\begin{equation}
T^{(2)}_{2\pi}
= 4\pi \int dr\, r^2 \psi^2_k(r)\, \widetilde{V}_{2\pi}^{(0)}(r) \, ,
\end{equation}
where $\psi_k(r)$ is the LO $\cs{1}{0}$ wave function. The ``outside'' part of $\psi_k(r)$ ($\Lambda^{-1} \lesssim r$) is subject to the Yukawa amplitude and is dominated by the irregular solution $\mathcal{H}_k(r)$~\eqref{eqn_coulomb} at short distance, which diverges as $1/r$ near $r \sim \Lambda^{-1}$. As a consequence, the UV divergence of $T^{(2)}_{2\pi}$ is illustrated by the integration of the outside wave function from any infrared length down to $r \sim \Lambda^{-1}$,
\begin{equation}
\begin{split}
T^{(2)}_{2\pi} &= 4\pi \int_{\sim \Lambda^{-1}} dr\, r^2 \psi^2_k(r)\, \widetilde{V}_{2\pi}^{(0)}(r) + \text{F.T.} \\
&\propto \left(\frac{\mathcal{N}}{C^{(0)}}\right)^2 \frac{{\chi_k^R}^2}{\left(C_R^{-1} - I_k^R\right)^2} \left( \rho_0 \Lambda^4 + \rho_1 k^2 \Lambda^2 + \rho_2 k^4 \ln \Lambda \right) + \text{F.T.} \, , \label{eqn_T2pi}
\end{split}
\end{equation} 
where $\mathcal{N}/C^{(0)}$ is RG invariant (see Appendix~\ref{app_wf}) and ``F.T.'' refers to finite terms. $\rho_i$ are functions of $\kappa_\pi/\Lambda$ and have at most logarithmic dependence on $\Lambda$. While $\rho_0 \Lambda^4$ and $\rho_1 k^2 \Lambda^2$ can be respectively subtracted by $C^{(2)}$~\eqref{eqn_Adef} and $D^{(1)}$~\eqref{eqn_Bdef}, the divergence proportional to $\rho_2 k^4 \ln \Lambda$ needs $E^{(0)}$~\eqref{eqn_T2E} to cancel. That is, the $\cs{1}{0}$ distorted-wave counterterm of TPE0--- the $E$ term--- has two more derivatives than the primordial counterterm--- the $D$ term.

Using the fact that a multiple-pion exchange with $\mathcal{O}(Q^n/\mhi^n)$ correction to TPE0 behaves as $1/r^{5+n}$ at short distance, and repeating the above procedure, we can eventually conclude that, for any multiple-pion exchange, the $\cs{1}{0}$ distorted-wave counterterm is a momentum polynomial with two more powers than its primordial counterpart.

We have seen two motivations to promote the $E$ counterterm to $\mathcal{O}(Q^2)$: (i) to control the regularization-scheme dependence of two insertions of the $D$ term and (ii) to absorb the distorted-wave UV divergences of two-pion exchanges.

Unfortunately, the integral in Eq.~\eqref{eqn_T2pi} cannot be evaluated analytically even as $\Lambda \to \infty$; thus we cannot express the full $\mathcal{O}(Q^2)$ amplitude in terms of the previously defined renormalized building blocks. But the structure of $V_S$ at $\mathcal{O}(Q^2)$ will suffice in the numerical calculations carried out later:
\begin{equation}
\langle \cs{1}{0} | V_S^{(2)} | \cs{1}{0} \rangle = C^{(2)} + \frac{D^{(1)}}{2}({p'}^2 + p^2) + E^{(0)}{p'}^2p^2 \, .\label{eqn_VS21p1}
\end{equation} 

\subsection{$P$ wave}

The distorted-wave counterterm for TPE0 in $\cp{1}{1}$ is the same as the primordial counterterm because, without an irregular component, the LO $P$-wave outside solution cannot make the distorted-wave counterterm more singular than the primordial one. It follows from this, combined with the observation that the residual counterterm for $\cp{1}{1}$ is not larger than TPE0, that WPC does not need to change for $\cp{1}{1}$; a single $P$-wave counterterm is what is needed for $\mathcal{O}(Q^2)$ and $\mathcal{O}(Q^3)$:
\begin{equation}
\langle \cp{1}{1} | V_S^{(2,\, 3)} | \cp{1}{1} \rangle = C^{(0,\, 1)}_\cp{1}{1} p' p \, .
\end{equation}

\subsection{Numerics\label{sec_NNLO_num}}

We compare our EFT calculations with the Nijmegen partial wave analysis (PWA)~\cite{Stoks:1993tb}. The expressions for the delta-less TPEs from Ref.~\cite{Epelbaum:1998ka-1999dj} are adopted here. Sharp momentum cutoff is used in solving the (partial-wave) Lippmann-Schwinger equation for the LO amplitudes and in evaluating the integrals involved in perturbative insertions of the subleading potentials. The analytical expressions of insertions of subleading counterterms, Eqs.~\eqref{eqn_T1B} and \eqref{eqn_T2S}, are not used since they are exactly correct only as $\Lambda \to \infty$.

\begin{figure}
\includegraphics[clip=true, scale=0.8]{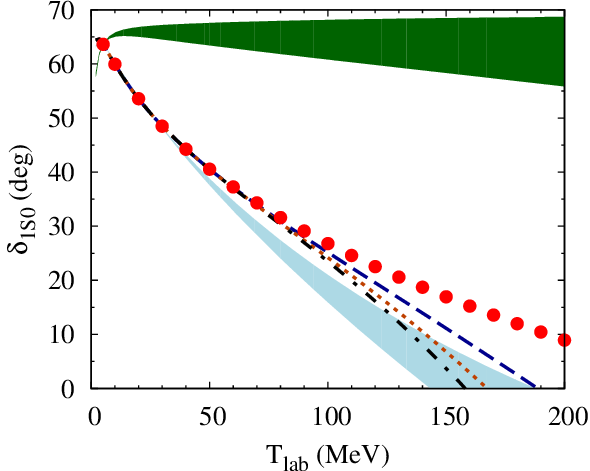}
\caption{(Color online) $\cs{1}{0}$ phase shifts as a function of laboratory energy. The red dots are from the Nijmegen PWA~\cite{Stoks:1993tb}. The dark green (light blue) band is the LO ($\mathcal{O}(Q)$) EFT result with $\Lambda = 0.5 - 2$ GeV. The dashed, dotted, and dot-dashed lines are $\mathcal{O}(Q^2)$ with $\Lambda = 0.5$, $1$, and $2$ GeV, respectively.\label{fig_tlab_1s0}}
\end{figure}

Plotted in Fig.~\ref{fig_tlab_1s0} are $\cs{1}{0}$ phase shifts versus laboratory energy, $T_\text{lab}$. The LO curve is fitted to the PWA at $T_\text{lab} = 5$ MeV. The PWA points at $T_\text{lab} = 25$ and 50 MeV are added to determine $D$ and $E$, respectively, at $\mathcal{O}(Q)$ and $\mathcal{O}(Q^2)$. A good reproduction of the PWA is achieved up to $T_\text{lab} \simeq 100$ MeV, which translates into $k \simeq 200$ MeV. 

Unlike in the triplet channels, the analytical arguments for renormalizability in the singlet channels are quite solid. So it is less crucial to examine numerically the cutoff (in)dependence of the EFT amplitudes. However, it is still reassuring to see that the $\mathcal{O}(Q^2)$ curve with $\Lambda = 1$ GeV is closer to the $\Lambda = 2$ GeV curve, suggesting the cutoff independence for large $\Lambda$.

\begin{figure}
\includegraphics[clip=true, scale=0.8]{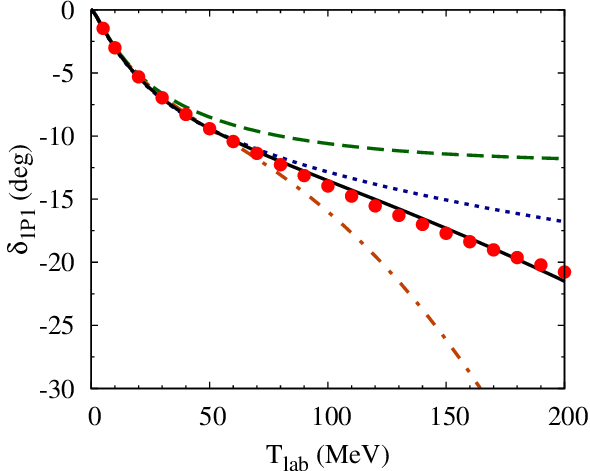}
\caption{(Color online) $\cp{1}{1}$ phase shifts as a function of laboratory energy. The red dots are from the PWA~\cite{Stoks:1993tb}. The dashed and dotted lines are respectively the LO and $\mathcal{O}(Q^2)$ EFT curves. The dot-dashed (set I) and solid (set II) lines are $\mathcal{O}(Q^3)$ curves with different sets of $\pi \pi NN$ seagull couplings (see the text for explanation).\label{fig_tlab_1p1}}
\end{figure}

Although WPC is intact for $\cp{1}{1}$, we plot in Fig.~\ref{fig_tlab_1p1} $\cp{1}{1}$ phase shifts, for completeness. The cutoff independence is rather trivial for $\cp{1}{1}$; therefore, only $\Lambda = 1.5$ GeV is employed. Since going to $\mathcal{O}(Q^3)$ in $\cp{1}{1}$ is much easier than in $\cs{1}{0}$, we include $\mathcal{O}(Q^3)$ results as well, in which the subleading TPE (TPE1) contributes. There is only one counterterm up to $\mathcal{O}(Q^3)$, $C_\cp{1}{1}$~\eqref{eqn_VS21p1}, which we determine by fitting to the PWA point at $T_\text{lab} = 50$ MeV.

TPE1 has crucial dependences on the $\pi \pi NN$ seagull couplings, $c_i$, that have chiral index $\nu = 1$. We show $\mathcal{O}(Q^3)$ EFT curves, respectively, with two commonly used sets of $c_i$ (in unit of GeV$^{-1}$): (I) the dot-dashed line with $c_{1}=-0.81$, $c_{3}=-4.7$, and $c_{4}=3.4$~\cite{Bu00} and (II) the solid line with $c_{1}=-0.81$, $c_{3}=-3.4$, and $c_{4}=3.4$~\cite{Entem:2002sf,Valderrama:2011mv}. The impact of the uncertainties of $c_i$ is significant beyond approximately $50$ MeV. Since the uncertainties of $c_i$ have their roots in slow convergence of the delta-less description of $\pi N$ scattering, we expect that the delta-ful pion exchanges~\cite{Ordonez:1993-1995, Kaiser:1998wa, Krebs:2007rh}, with the $\pi N \Delta$ low-energy constants determined by the $\pi N$ scattering data around the delta peak~\cite{Pascalutsa:2002pi, Long:2009wq, Long:2010kt}, will improve the convergence of chiral $NN$ EFT. In fact, aside from the open issues of power counting counterterms, the delta-ful nuclear forces have been shown to achieve a more rapid convergence in the two-nucleon~\cite{Valderrama:2008kj_2010fb, Entem:2007jg} and, on a more qualitative level, in the three-nucleon~\cite{Pandharipande:2005sx} sectors.

In the light of findings of Ref.~\cite{Baru:2012iv}, a few more remarks about our numerical results are in order. Reference~\cite{Baru:2012iv} argues that higher terms of a certain series of irreducible multiple-pion exchanges, the multiple-scattering series (MSS), are suppressed by a mass scale ($M_\text{MSS}$) much smaller than $4\pi f_\pi \sim 1.2$ GeV, which was estimated by WPC. Using the conversion of coordinate cutoff to momentum cutoff~\cite{Entem:2007jg}, $\Lambda = \pi/(2\mathcal{R}_c)$, we translate the breakdown length scale of the MSS found in Ref.~\cite{Baru:2012iv} into $M_\text{MSS} \simeq 390 - 620$ MeV, depending on the value of $c_3$. Assuming the finding in Ref.~\cite{Baru:2012iv} to be correct, $M_\text{MSS}$, instead of $4\pi f_\pi$, may now be the breakdown scale of chiral EFT itself, and one may then be able to choose $\Lambda$ with a value as low as $M_\text{MSS}$. But there seems to be nothing wrong with choosing a cutoff value that is higher than the actual breakdown scale, provided that subleading orders are treated in perturbation theory. We at least already know that this is the case for the pionless EFT and single-nucleon chiral perturbation theory. What is more important is whether the slower-than-expected convergence is reflected at the level of on-shell amplitudes. Interestingly, the $NN$ phase shifts calculated with delta-less TPEs and with our power counting, shown in Ref.~\cite{Long:2011qx} and in this paper, indeed suggest a breakdown scale comparable to or even lower than $M_\text{MSS}$, for which the slow convergence of the MSS in the delta-less theory may be suspected as the culprit.

\section{Discussion and Conclusion\label{sec_discussion}}

We have studied how RG invariance constrains in the singlet channels of $NN$ scattering the structure of subleading counterterms, with $S$ and $P$ waves as the examples. Our analysis shows a hierarchy of $\cs{1}{0}$ counterterms that resembles the pionless theory while WPC remains unchanged in $\cp{1}{1}$ and higher singlet partial waves.

To facilitate the discussion, the subleading counterterms are classified into three categories according to the loop diagrams that drive their evolution (see Sec.~\ref{sec_SLC}). The residual counterterms eliminate the small cutoff dependence of the LO amplitude, in order to achieve the exact RG invariance. The primordial and distorted-wave counterterms are the short-range operators necessary to absorb the divergences of multiple-pion exchanges sandwiched between free and LO interacting states, respectively.

We have argued that RG invariance provides two mechanisms to enhance, relative to WPC, the short-range forces in $\cs{1}{0}$. (i) As the residual counterterms for the LO amplitude, they scale similarly to the contact interactions of the pionless theory~\cite{bira-pionless, Kaplan:1998tg}: for a generic $\cs{1}{0}$ counterterm $C_{2n}$ with $2n$ derivatives, 
\begin{equation}
C^\text{res}_{2n} \sim \frac{4\pi}{m_N} \frac{1}{\mlo^{n+1} \mhi^n} \, . \label{eqn_c2nres}
\end{equation} 
(ii) As the distorted-wave counterterms for multiple-pion exchanges, they are enhanced by $\mathcal{O}(\mhi^2/\mlo^2)$, but only starting from the four-derivative term,
\begin{equation}
C^\text{dis}_{2n} \sim \frac{4\pi}{m_N} \frac{1}{\mlo^3 \mhi^{2n-2}} \, , \qquad n \geqslant 2 \, .
\end{equation}
Since the enhancement due to the residual counterterms dominates, we power count $\cs{1}{0}$ contact interaction according to Eq. \eqref{eqn_c2nres}, as if the theory were the pionless one. In particular, the $\mathcal{O}(Q^3)$ counterterms have the following structure:
\begin{equation}
\langle \cs{1}{0} | V_S^{(3)} | \cs{1}{0} \rangle = C^{(3)} + \frac{D^{(2)}}{2}\left({p'}^2 + p^2\right) + E^{(1)}{p'}^2 p^2 + \frac{F^{(0)}}{2} {p'}^2 p^2\left({p'}^2 + p^2\right) \, .
\end{equation} 
The numerical implementation of $\mathcal{O}(Q^3)$ for $\cs{1}{0}$ is currently being worked on and will be reported in later publications. Summarized in Table~\ref{table:PC} is our power counting for the two-nucleon sector in both singlet and triplet channels for $S$ and $P$ waves.

\begin{table}[bt]
\begin{tabular}{|c|c|}
  \hline
  $\mathcal{O}(1)$ & OPE,\; $C_\cs{1}{0}$,\;
  $\begin{pmatrix} C_\cs{3}{1} & 0 \\
  0 & 0
  \end{pmatrix} $,\;
  $C_\cp{3}{0}p' p$,\;
  $\begin{pmatrix} C_\cp{3}{2} p' p & 0 \\
  0 & 0
  \end{pmatrix}$ \\
  \hline
  $\mathcal{O}(Q)$ & $D_\cs{1}{0}({p'}^2 + p^2)$ \\
  \hline
  $\mathcal{O}(Q^2)$ & TPE0,\; $E_\cs{1}{0}{p'}^2 p^2$,\;
  $\begin{pmatrix} D_\cs{3}{1}({p'}^2 + p^2) & E_\text{SD}\,p^2 \\
  E_\text{SD}\,{p'}^2 & 0
  \end{pmatrix} $,\; \\
  & $D_\cp{3}{0}\,p' p({p'}^2 + p^2)$,\;
  $p' p \begin{pmatrix} D_\cp{3}{2}({p'}^2 + p^2) & E_\text{PF}\,p^2 \\
  E_\text{PF}\,{p'}^2 & 0
  \end{pmatrix} $, \\
  & $C_\cp{1}{1} p' p$,\; $C_\cp{3}{1}p' p$ \\
  \hline
  $\mathcal{O}(Q^3)$ & TPE1,\; $F_\cs{1}{0}{p'}^2p^2({p'}^2 + p^2)$ \\
  \hline
\end{tabular}
\caption{Power counting for pion exchanges, $S$ and $P$-wave counterterms up to $\mathcal{O}(Q^3)$. $p$ ($p'$) is the magnitude of the center-of-mass incoming (outgoing) momentum. The two-by-two matrices are for the coupled channels.\label{table:PC}}
\end{table}

We scrutinize WPC with a more stringent interpretation of RG invariance: not only should the cutoff dependence become vanishingly small for $\Lambda \gtrsim \mhi$, but it must vanish sufficiently fast so that the accuracy claimed by the power counting is consistent with the cutoff error. This leads to a crucial conclusion in our analysis that, contrary to WPC, $\mathcal{O}(Q)$ of the EFT expansion does not vanish. Instead, $\mathcal{O}(Q)$ is made of one insertion of the two-derivative $\cs{1}{0}$ counterterm: $D/2({p'}^2+p^2)$. Although we are not the first to propose this, our argument, that the cutoff error of the LO amplitude is one order lower than TPE0 and has to be corrected by the $D$ term alone, provides some new insights. For instance, unlike Ref.~\cite{Kaplan:1996xu} (also discussed later in Ref.~\cite{Gegelia}), our rationale is \textit{a priori} and does not rely on the numerical value of $D$ in a particular renormalization scheme.

A full, nonperturbative RG analysis, with OPE as the only long-range force, of the counterterms was attempted in Refs.~\cite{Barford:2002je, Birse:2005um}, in which it was also concluded that the $D$ counterterm is more important than TPE0. Although the nonperturbative RG analysis appears to be free of any guesswork for obtaining power counting, the robustness of the conclusions of Refs.~\cite{Barford:2002je, Birse:2005um} is obscured by the assumptions made therein to derive and solve the RG equation. On the other hand, our approach can be viewed as the explicit, order-by-order examination of an ansatz--- the proposed power counting--- to the RG equation. If RG invariance can be shown to hold at all orders, which we could not rigorously achieve though, we cannot think of any reason why the proposed power counting could not be \emph{one} of the solutions to the RG equation. In other words, we think that there may be more than one RG-invariant power counting, and only the data or the underlying theory can tell which one is more efficient.

It is instructive to compare the power counting of $\cs{1}{0}$ with that of the attractive triplet channels. A nonvanishing $\mathcal{O}(Q)$ arising in $\cs{1}{0}$ but not in the triplet channels has everything to do with the fact that OPE is regular ($1/r$) in $\cs{1}{0}$ but singular ($1/r^3$) in the triplet channels. It is interesting that the singular attraction of OPE costs a few more LO counterterms in the attractive triplet channels (e.g., $\cp{3}{0}$ and $\cpf$) but in the meantime it avoids the pionless-theory-like proliferation of subleading counterterms.

The distorted-wave enhancement to the singlet-channel short-range forces occurs in only $S$ wave ($\cs{1}{0}$), and it affects the power counting to a lesser extent than that of the residual counterterms. In contrast, the distorted-wave enhancement in the attractive triplet channels takes place in higher partial waves ($\cp{3}{0}$, $\cpf$, etc.) but not in $S$ wave, and it plays more important role in power counting than the residual counterterms.

\acknowledgments We thank Bira van Kolck and Daniel Phillips for their encouragement and critical discussions on the topic, and Martin Savage for reminding us of the $m_\pi$ dependence of the leading counterterm. We are grateful for hospitality to the National Institute for Nuclear Theory (INT) at the University of Washington and the organizers of the INT program ``Simulations and Symmetries: Cold Atoms, QCD, and Few-hadron Systems,'' in which the work was stimulated. B.w.L. thanks the nuclear theory group at the George Washington University and the TQHN group at the University of Maryland for their hospitality and Harald Greisshammer and Paulo Bedaque for useful discussions. C.J.Y. thanks B.~Barrett for his valuable support. This work is supported by the U.S. DOE under Contracts No.DE-AC05-06OR23177 (B.w.L.) and No. DE-FG02-04ER41338 (C.J.Y.), and by the NSF under Grant No. PHYS-0854912 (C.J.Y.), and is coauthored by Jefferson Science Associates, LLC under U.S. DOE Contract No. DE-AC05-06OR23177.

\appendix

\section{LO wave function\label{app_wf}}

With the \emph{regularized} LO potential, the $S$-wave radial wave function is well defined at the origin and can be written as
\begin{equation}
\psi_k(r) = \psi_k(0) \phi_k(r) \, ,
\label{eqn_psikphik}
\end{equation}
where $\phi_k(r)$ is the regular solution in the sense $\phi_k(r) \to j_0(kr)$ as $r \to 0$, with $j_0(\rho)$ being the zeroth spherical Bessel function.

With regularization, $\widetilde{V}_Y(r)$--- the Fourier transform of $V_Y(q)$--- becomes relatively flat on the inside while it resumes the Yukawa form on the outside. The LO contact potential, $V_S^{(0)}$, is smeared inside and vanishes outside. This means that the inside wave function is largely decided by $C^{(0)}(\Lambda)$ and $\Lambda$, whereas the outside part is dominated by a combination of the irregular ($\mathcal{H}_k(r)$) and regular ($\mathcal{J}_k(r)$) solutions to the Yukawa potential,
\begin{equation}
\phi_k(r) = \mathcal{N}(C^{(0)}, \Lambda)\left[\mathcal{H}_k(r) + \theta(C^{(0)}, \Lambda)\mathcal{J}_k(r)\right]\, , \qquad r \gtrsim \Lambda^{-1} \, .\label{eqn_phiout}
\end{equation}
$\mathcal{H}_k(r)$ and $\mathcal{J}_k(r)$ have the following small $kr$ expansions:
\begin{equation}
\begin{split}
\mathcal{J}_k(r) &= \sum_{n=0}\xi_n\left(\kappa_\pi r\right) (kr)^{2n} \, , \\
\mathcal{H}_k(r) &= \frac{1}{\kappa_\pi r} \sum_{n=0} \delta_n\left(\kappa_\pi r\right) (kr)^{2n} - 2 \mathcal{J}_k(r) \ln \left(\kappa_\pi r\right) \, ,
\end{split}\label{eqn_coulomb}
\end{equation} 
where $\xi_n(x)$ and $\delta_n(x)$ are analytic functions around $x = 0$ and can be further expanded to obtain the expansions of $\mathcal{J}_k(r)$ and $\mathcal{H}_k(r)$ in powers of $r$. $\mathcal{N}$ and $\theta$ in Eq.~\eqref{eqn_phiout} are functions of $C^{(0)}(\Lambda)$ and $\Lambda$ because the inside and outside wave functions need to match near $r \sim \Lambda^{-1}$ when $k = 0$ or any other small momentum for which we decide to fit $C^{(0)}(\Lambda)$.

On the other hand, the three-dimensional (in-state) wave function is related to the LO $T$-matrix by
\begin{equation}
\psi_{\vec{k}} (\vec{x}\,) = e^{i\vec{k}\cdot\vec{x}} + \int \frac{d^3l}{(2\pi)^3} e^{i\vec{l}\cdot\vec{x}} \frac{T^{(0)}(\vec{l},\vec{k};k)}{E - \frac{l^2}{m_N} + i\epsilon} \, . \label{eqn_psikx}
\end{equation}
Therefore, $\psi_k(0)$ is given by 
\begin{equation}
\psi_k(0) = 1 + \int \frac{d^3l}{(2\pi)^3} \frac{T^{(0)}(\vec{l},\vec{k}; k)}{E - \frac{l^2}{m_N} + i\epsilon} \, . \label{eqn_fk}
\end{equation}
Since $T^{(0)}(\vec{l},\vec{k}; k)$ is generated by the regularized LO potential $V^{(0)}(\vec{p}\,', \vec{p}\,) \mathcal{F}_R(\vec{p}\,'/\Lambda, \vec{p}/\Lambda)$, it dies off in the UV region. So an additional regularization of the integrals in the above equations is unnecessary. $\psi_{\vec{k}} (\vec{x}\,)$ satisfies the Schr\"odinger equation,
\begin{equation}
-\vec{\nabla}^2 \psi_{\vec{k}}(\vec{x}\,) + m_N \int d^3x' \widetilde{V}^{(0)}_\Lambda(\vec{x}, \vec{x}\,') \psi_{\vec{k}}(\vec{x}\,') = k^2 \psi_{\vec{k}}(\vec{x}\,) \, ,
\label{eqn_intSch}
\end{equation} 
where the regularized LO potential is generally nonlocal at short distance, $r \sim \Lambda^{-1}$,
\begin{equation}
\widetilde{V}^{(0)}_\Lambda(\vec{x}, \vec{x}\,') = \int \frac{d^3l}{(2\pi)^3} \frac{d^3l'}{(2\pi)^3} \mathcal{F}_R\left(\frac{\vec{l}}{\Lambda}\, , \frac{\vec{l}\,'}{\Lambda} \right) V^{(0)}\left(|\vec{l}-\vec{l}\,'| \right) e^{i (\vec{l}\cdot\vec{x} - \vec{l}\,'\cdot\vec{x}\,')} \, ,
\end{equation}
unless the cutoff regulator depends only on the momentum transfer.
In the limit $\Lambda \to \infty$, the nonlocal effect disappears and the Schr\"odinger equation becomes local but formal with the un-regularized LO potential:
\begin{equation}
-\vec{\nabla}^2 \psi_{\vec{k}}(\vec{x}\,) + m_N \widetilde{V}^{(0)}(\vec{x}\,) \psi_{\vec{k}}(\vec{x}\,) = k^2 \psi_{\vec{k}}(\vec{x}\,) \, ,
\label{eqn_bareScheqn}
\end{equation} 
where
\begin{equation}
\widetilde{V}^{(0)}(\vec{x}) = C^{(0)}\delta^{(3)}(\vec{x}) - \frac{4\pi \kappa_\pi}{m_N} \frac{e^{-m_\pi r}}{r} \, .\label{eqn_VLOtilde}
\end{equation} 
One could use a regulator that depends only on the momentum transfer so that the Schr\"odinger equation becomes exactly local even for finite $\Lambda$. But with such a regulator it is difficult to resum $C^{(0)}$ analytically because the bubble diagrams--- iterations of only $C^{(0)}$--- no longer form a geometrical series. 

If $\mathcal{F}_R(\vec{x}, \vec{y}\,)$ is separable, $\mathcal{F}_R(\vec{x}, \vec{y}\,) = f_R(|\vec{x}|) f_R(|\vec{y}|)$, the analytical results for LO in Sec.~\ref{sec_LOandNLO} are exact for finite $\Lambda$. With such a regulator, we can use Eqs.~\eqref{eqn_TLO}, \eqref{eqn_chiE}, and \eqref{eqn_IE} to obtain
\begin{equation}
\psi_k(0) = \frac{1}{C^{(0)}} \frac{\chi(k; k)}{{(C^{(0)})}^{-1} - I_k} \, . \label{eqn_fkchiE}
\end{equation}
Using Eq.~\eqref{eqn_psikphik}, we find the asymptotic form of $\psi_k(r)$, which must be RG invariant in order to extract scattering observables,
\begin{equation}
\psi_k(r) = \frac{\mathcal{N}}{C^{(0)}} \frac{\chi(k; k)}{(C^{(0)})^{-1} - I_k} \left[\mathcal{H}_k(r) + \theta(C^{(0)}, \Lambda)\mathcal{J}_k(r)\right] \, , \qquad r \gg \Lambda^{-1}\, .
\end{equation} 
Recalling that $\chi(k; k)/[(C^{(0)})^{-1} - I_k] \to \chi^R(k; k)/(C_R^{-1} - I_k^R)$, one concludes that $\mathcal{N}/C^{(0)}$ and $\theta$ are both RG invariant as $\Lambda \to \infty$.

\section{Useful integrals for subleading $T$-matrix\label{app_tricks}}

We briefly describe the computations of the integrals stemming from insertions of subleading counterterms into the LO $T$-matrix, such as Eqs.\eqref{eqn_T1B}, \eqref{eqn_T2S}, \eqref{eqn_T2E}, and \eqref{eqn_T2epsilon}. Note that the results shown here are only formal expressions for a generic regulator in the large $\Lambda$-limit.

When evaluating $(1 + T^{(0)}G) V_S^{(1)} (G T^{(0)} + 1)$~\eqref{eqn_T1B}, we need to know the following integral:
\begin{equation}
\Sigma_2(k) \equiv k^2 + \int \frac{d^3l}{(2\pi)^3} l^2 \frac{T^{(0)}(\vec{l},\vec{k}; k)}{E - \frac{l^2}{m_N} + i\epsilon} \, .
\end{equation}
By differentiating with respect to $\vec{x}$ on both sides of Eq.~\eqref{eqn_psikx} and letting $\vec{x} = 0$ in the end, one finds
\begin{equation}
\Sigma_2(k) = -\vec{\nabla}^2 \psi_{k}(0)\, .
\end{equation}
For finite $\Lambda$, $\vec{\nabla}^2 \psi_{k}(0)$ does not enjoy a simple relation to quantities at the origin, since ${\widetilde{V}}^{(0)}_\Lambda(\vec{x}, \vec{x}\,')$ is generally nonlocal at short distance, as indicated in Eq.~\eqref{eqn_intSch}. But as $\Lambda \to \infty$, one can use Eqs.~\eqref{eqn_bareScheqn} and \eqref{eqn_fkchiE} to obtain
\begin{equation}
\Sigma_2(k) = \frac{1}{C^{(0)}} \left[k^2 - m_N \widetilde{V}^{(0)}(0)\right] \frac{\chi_k^R}{C_R^{-1} - I_k^R} \qquad \text{at}\;\; \Lambda \to \infty \, .
\end{equation}  
More generally,
\begin{equation}
\Sigma_{2n}(k) \equiv k^{2n} + \int \frac{d^3l}{(2\pi)^3} l^{2n} \frac{T^{(0)}(\vec{l},\vec{k}; k)}{E - \frac{l^2}{m_N} + i\epsilon} = (-1)^n \vec{\nabla}^{2n} \psi_k(0) \, ,
\end{equation}
where $\vec{\nabla}^{2n} \psi_k(0)$ can be related by successive differentiation on Eq.~\eqref{eqn_bareScheqn} to $\widetilde{V}^{(0)}(\vec{x}\,)$ and its derivatives at $r = 0$.

Integrals involving the LO interacting Green function $\mathscr{G}_k$ are encountered in computing Eq.~\eqref{eqn_T2S}:
\begin{equation}
\Pi_{n, m}(k) \equiv \int \frac{d^3 l_1}{(2\pi)^3} \frac{d^3 l_2}{(2\pi)^3}\, l_1^{2n} l_2^{2m}\, \mathscr{G}_k(\vec{l}_2, \vec{l}_1) \, , \label{eqn_intlm}
\end{equation} 
with
\begin{equation}
\mathscr{G}_k(\vec{l}_2, \vec{l}_1) \equiv (2\pi)^3 \frac{\delta^{(3)}(\vec{l}_1 - \vec{l}_2)}{E - \frac{l_1^2}{m_N} + i\epsilon} + \frac{T^{(0)}(\vec{l}_2, \vec{l}_1; k)}{(E - \frac{l_2^2}{m_N} + i\epsilon)(E - \frac{l_1^2}{m_N} + i\epsilon)} \, .
\end{equation}
The generating function for these integrals is the coordinate-space version of $\mathscr{G}_k$,
\begin{equation}
\widetilde{\mathscr{G}}_k(\vec{x}_2, \vec{x}_1) = \int \frac{d^3 l_1}{(2\pi)^3} \frac{d^3 l_2}{(2\pi)^3} \mathscr{G}_k(\vec{l}_2, \vec{l}_1) e^{i(\vec{l}_2\cdot\vec{x}_2 - \vec{l}_1\cdot\vec{x}_1)} \, ,
\end{equation}
and
\begin{equation}
\Pi_{n, m}(k) = (-1)^{n+m} \vec{\nabla}^{2n}_{x_1} \vec{\nabla}^{2m}_{x_2} \widetilde{\mathscr{G}}_k(\vec{x}_2, \vec{x}_1) |_{\vec{x}_1 = 0,\, \vec{x}_2 = 0} \, .
\end{equation} 
The second derivative of $\widetilde{\mathscr{G}}_k(\vec{x}_2, \vec{x}_1)$ is given by
\begin{equation}
-\vec{\nabla}^2_{\{2,1\}} \widetilde{\mathscr{G}}_k(\vec{x}_2, \vec{x}_1) = -m_N \delta^{(3)}(\vec{x}_2 - \vec{x}_1) + \left[k^2 - m_N \widetilde{V}^{(0)}(\vec{x}_{\{2, 1\}}\,)\right] \widetilde{\mathscr{G}}_k(\vec{x}_2, \vec{x}_1) \, . \label{eqn_gk2d}
\end{equation} 
Using Eqs.~\eqref{eqn_TLO}, \eqref{eqn_chiE}, \eqref{eqn_IE}, and \eqref{eqn_CRdef}, we can write $\widetilde{\mathscr{G}}_k(0, 0)$ as
\begin{equation}
\widetilde{\mathscr{G}}_k(0, 0) = I_k + \frac{I_k^2}{(C^{(0)})^{-1} - I_k} = \frac{1/(C^{(0)})^2}{C_R^{-1} - I_k^R} - 1/C^{(0)} \,  \qquad \text{for}\;\; \Lambda \to \infty \, .
\end{equation} 
Again, successive differentiation with respect to $\vec{x}_{1, 2}$ on both sides of Eq.~\eqref{eqn_gk2d} and letting $\vec{x}_{1, 2} = 0$ leads to $\Pi_{n, m}(k)$ with larger $m$ and/or $n$.


\begin{thebibliography}{99}

\bibitem{Weinberg:1990-1991} S.~Weinberg, 
%``Nuclear forces from chiral Lagrangians,''
Phys.\ Lett.\ \textbf{B251}, 288 (1990);
%``Effective Chiral Lagrangians For Nucleon - Pion Interactions And Nuclear Forces,''
Nucl.\ Phys.\ \textbf{B363}, 3 (1991). %%CITATION = NUPHA,B363,3;%%

\bibitem{Ordonez:1993-1995} C.~Ordonez, L.~Ray, and U.~van Kolck, 
%``Nucleon-Nucleon Potential From An Effective Chiral Lagrangian,''
Phys.\ Rev.\ Lett.\ \textbf{72}, 1982 (1994); %%CITATION = PRLTA,72,1982;%%
%  C.~Ordonez, L.~Ray and U.~van Kolck,
%``The Two-Nucleon Potential from Chiral Lagrangians,''
Phys.\ Rev.\ \textbf{C53}, 2086 (1996). %  [arXiv:hep-ph/9511380].
%%CITATION = PHRVA,C53,2086;%%

\bibitem{Epelbaum:1998ka-1999dj} E.~Epelbaum, W.~Gloeckle, and U.~-G.~Meissner, 
%``Nuclear forces from chiral Lagrangians using the method of unitary transformation. 1. Formalism,''
Nucl.\ Phys.\ \textbf{A637}, 107 (1998); %  [nucl-th/9801064];
%\cite{Epelbaum:1999dj}
%\bibitem{Epelbaum:1999dj}
%  E.~Epelbaum, W.~Gloeckle, U.~-G.~Meissner,
%``Nuclear forces from chiral Lagrangians using the method of unitary transformation. 2. The two nucleon system,''
Nucl.\ Phys.\ \textbf{A671}, 295 (2000). %[nucl-th/9910064].

\bibitem{Epelbaum:2004fk} E.~Epelbaum, W.~Glockle, and U.~G.~Meissner, 
%``The two-nucleon system at next-to-next-to-next-to-leading order,''
Nucl.\ Phys.\ \textbf{A747}, 362 (2005).
% [arXiv:nucl-th/0405048]. 
%%CITATION = NUPHA,A747,362;%%

\bibitem{Entem:2001cg} D.~R.~Entem and R.~Machleidt, 
%``Accurate nucleon-nucleon potential based upon chiral perturbation theory,''
Phys.\ Lett.\ \textbf{B524}, 93 (2002). %  [nucl-th/0108057].

\bibitem{Entem:2002sf}
D.~R.~Entem and R.~Machleidt,
%``Chiral 2pi exchange at order four and peripheral NN scattering,''
Phys.\ Rev.\ \textbf{C66}, 014002 (2002). %  [nucl-th/0202039].

\bibitem{Kaplan:1996xu} D.~B.~Kaplan, M.~J.~Savage, and M.~B.~Wise, 
%``Nucleon - nucleon scattering from effective field theory,''
Nucl.\ Phys.\ \textbf{B478}, 629 (1996).

\bibitem{Beane:2000wh} S.~R.~Beane, P.~F.~Bedaque, L.~Childress,
A.~Kryjevski, J.~McGuire, and U.~van Kolck, 
%``Singular Potentials and Limit Cycles,''
Phys.\ Rev.\ \textbf{A64}, 042103 (2001). %[arXiv:quant-ph/0010073].
%%CITATION = PHRVA,A64,042103;%%

\bibitem{Beane:2001bc} S.~R.~Beane, P.~F.~Bedaque, M.~J.~Savage, and U.~van
Kolck, %``Towards a perturbative theory of nuclear forces,''
Nucl.\ Phys.\ \textbf{A700}, 377 (2002). %%CITATION = NUPHA,A700,377;%%

\bibitem{Nogga:2005hy} A.~Nogga, R.~G.~E.~Timmermans, and U.~van Kolck, 
%``Renormalization of One-Pion Exchange and Power Counting,''
Phys.\ Rev.\ \textbf{C72}, 054006 (2005). %%CITATION = PHRVA,C72,054006;%%

\bibitem{Birse:2005um} M.~C.~Birse, 
%``Power counting with one-pion exchange,''
Phys.\ Rev.\ \textbf{C74}, 014003 (2006).% [nucl-th/0507077].

\bibitem{saopaulo} T.~Frederico, V.~S.~Timoteo, and L.~Tomio, 
%``Renormalization of the one pion exchange interaction,''
Nucl.\ Phys.\ \textbf{A653}, 209 (1999);
V.~S.~Timoteo, T.~Frederico, A.~Delfino, and L.~Tomio, 
Phys.\ Lett.\ \textbf{B621}, 109 (2005);
%V.~S.~Timoteo, T.~Frederico, A.~Delfino and L.~Tomio,
Phys.\ Rev.\  {\bf C83}, 064005 (2011);
S.~Szpigel and V.~S.~Timoteo,
%``Power counting and renormalization group invariance in the subtracted kernel method for the two-nucleon system,''
e-print arXiv:1112.5972 [nucl-th].
%%CITATION = ARXIV:1112.5972;%%

\bibitem{YangOhio} C.~J.~Yang, C.~Elster, and D.~R.~Phillips, 
%``Subtractive renormalization of the chiral potentials up to next-to-next-to-leading order in higher NN partial waves,''
Phys.\ Rev.\ \textbf{C77}, 014002 (2008);
Phys.\ Rev.\ \textbf{C80}, 034002 (2009);
Phys.\ Rev.\ \textbf{C80}, 044002 (2009).

\bibitem{Valderrama:2009ei} M.~P.~Valderrama, 
%``Perturbative renormalizability of chiral two pion exchange in nucleon-nucleon scattering,''
Phys.\ Rev.\ \textbf{C83}, 024003 (2011).% [arXiv:0912.0699 [nucl-th]]

\bibitem{Valderrama:2011mv}
M.~P.~Valderrama,
%``Perturbative Renormalizability of Chiral Two Pion Exchange in Nucleon-Nucleon Scattering: P- and D-waves,''
Phys.\ Rev.\ {\bf C84}, 064002 (2011).
% [arXiv:1108.0872 [nucl-th]].

\bibitem{Long:2011qx} 
Bingwei~Long and C.~J.~Yang,
%``Renormalizing chiral nuclear forces: a case study of 3P0,''
Phys.\ Rev.\ {\bf C84}, 057001 (2011).
% [arXiv:1108.0985 [nucl-th]].
%%CITATION = ARXIV:1108.0985;%%

\bibitem{Long:2011xw} 
  Bingwei~Long and C.~J.~Yang,
  %``Renormalizing Chiral Nuclear Forces: Triplet Channels,''
  Phys.\ Rev.\ {\bf C85}, 034002 (2012).
%   [arXiv:1111.3993 [nucl-th]].
  %%CITATION = ARXIV:1111.3993;%%

\bibitem{Phillips:1996ae}
  D.~R.~Phillips and T.~D.~Cohen,
  %``How short is too short? Constraining contact interactions in nucleon-nucleon scattering,''
  Phys.\ Lett.\  {\bf B390}, 7 (1997).
%   [nucl-th/9607048].

\bibitem{PavonValderrama:2004nb}
  M.~Pavon Valderrama and E.~Ruiz Arriola,
  %``Renormalization of NN-scattering with one pion exchange and boundary conditions,''
  Phys.\ Rev.\  {\bf C70}, 044006 (2004).
%   [nucl-th/0405057].


\bibitem{Barford:2002je}
  T.~Barford and M.~C.~Birse,
  %``A Renormalization group approach to two-body scattering in the presence of long range forces,''
  Phys.\ Rev.\  {\bf C67}, 064006 (2003);
%   [hep-ph/0206146].
%\cite{Birse:1998dk}
  M.~C.~Birse, J.~A.~McGovern, and K.~G.~Richardson,
  %``A Renormalization group treatment of two-body scattering,''
  Phys.\ Lett.\ {\bf B464}, 169 (1999).
%   [hep-ph/9807302].
  %%CITATION = HEP-PH/9807302;%%


\bibitem{PavonValderrama:2007nu} M.~Pavon Valderrama and E.~R.~Arriola, 
%``Renormalization group analysis of boundary conditions in potential scattering,''
Annals Phys.\ (N.Y.) \textbf{323}, 1037 (2008).% [arXiv:0705.2952 [nucl-th]].

\bibitem{bira-private}
U.~van Kolck (private communication).

\bibitem{Kaplan:1998tg} D.~B.~Kaplan, M.~J.~Savage, and M.~B.~Wise, 
%%``A new expansion for nucleon nucleon interactions,''
Phys.\ Lett.\ \textbf{B424}, 390 (1998); %%CITATION = PHLTA,B424,390;%%
%\bibitem{Kaplan:1998we} D.~B.~Kaplan, M.~J.~Savage and M.~B.~Wise, 
%%``Two-nucleon systems from effective field theory,''
Nucl.\ Phys.\ \textbf{B534}, 329 (1998). %%CITATION = NUPHA,B534,329;%%

\bibitem{vanKolck:1999mw} U.~van Kolck, 
%``Effective field theory of nuclear forces,''
Prog.\ Part.\ Nucl.\ Phys.\ \textbf{43}, 337 (1999). 
%%CITATION = PPNPD,43,337;%%

\bibitem{Epelbaum:2006pt} 
  E.~Epelbaum and U.~-G.~Meissner,
  %``On the renormalization of the one-pion exchange potential and the consistency of Weinberg's power counting,''
e-print arXiv:nucl-th/0609037.
  %%CITATION = NUCL-TH/0609037;%%

\bibitem{Epelbaum:2009sd}
E.~Epelbaum and J.~Gegelia, 
%``Regularization, renormalization and 'peratization' in effective field
%theory for two nucleons,''
Eur.\ Phys.\ J.\ \textbf{A41}, 341 (2009). %[arXiv:0906.3822 [nucl-th]].
%%CITATION = EPHJA,A41,341;%%

\bibitem{Epelbaum:2012vx} 
  E.~Epelbaum and U.~-G.~Meissner,
  %``Chiral dynamics of few- and many-nucleon systems,''
  e-print arXiv:1201.2136 [nucl-th].
  %%CITATION = ARXIV:1201.2136;%%

\bibitem{Beane:2000fi}
S.~R.~Beane and M.~J.~Savage,
  %``Rearranging pionless effective field theory,''
  Nucl.\ Phys.\  {\bf A694}, 511 (2001).
%   [nucl-th/0011067].

\bibitem{Birse:2009my} M.~C.~Birse, 
%``More effective theory of nuclear forces,''
in \textit{6th International Workshop on chiral Dynamics,} PoS(CD09)078. %  [arXiv:0909.4641 [nucl-th]].

\bibitem{Entem:2009mf} 
  D.~R.~Entem and E.~Ruiz Arriola,
  %``Renormalization of the Off-shell chiral two-pion exchange NN interactions,''
  Phys.\ Rev.\ {\bf C80}, 047001 (2009).
%   [arXiv:0906.1945 [nucl-th]].
  %%CITATION = ARXIV:0906.1945;%%

\bibitem{Scaldeferri:1996nx} 
  K.~A.~Scaldeferri, D.~R.~Phillips, C.~W.~Kao, and T.~D.~Cohen,
  %``Short range interactions in an effective field theory approach for nucleon-nucleon scattering,''
  Phys.\ Rev.\ {\bf C56}, 679 (1997).
%   [nucl-th/9610049].
  %%CITATION = NUCL-TH/9610049;%%


\bibitem{Long:2007vp} Bingwei~Long and U.~van Kolck, 
%``Renormalization of Singular Potentials and Power Counting,''
Annals Phys.\ (N.Y.) \textbf{323}, 1304 (2008). %[arXiv:0707.4325 [quant-ph]].
%%CITATION = APNYA,323,1304;%%

\bibitem{Bedaque:2002mn} P.~F.~Bedaque and U.~van Kolck, 
%``Effective field theory for few-nucleon systems,''
Ann.\ Rev.\ Nucl.\ Part.\ Sci.\ \textbf{52}, 339 (2002). 
%[arXiv:nucl-th/0203055].
%%CITATION = ARNUA,52,339;%%

\bibitem{Fleming:1999ee} S.~Fleming, T.~Mehen, and I.~W.~Stewart, 
%``NNLO corrections to nucleon nucleon scattering and perturbative pions,''
Nucl.\ Phys.\ \textbf{A677}, 313 (2000).% [arXiv:nucl-th/9911001]. 
%%CITATION = NUPHA,A677,313;%%

\bibitem{Stoks:1993tb} V.~G.~J.~Stoks, R.~A.~M.~Klomp, M.~C.~M.~Rentmeester
and J.~J.~de Swart, 
%``Partial wave analysis of all nucleon-nucleon scattering data below
%350-MeV,''
Phys.\ Rev.\ \textbf{C48}, 792 (1993) (http://nn-online.org).

\bibitem{Bu00}
  P. Buttiker and U.~-G.~Meissner,
  Nucl.\ Phys.\ {\bf A668}, 97 (2000).

\bibitem{Kaiser:1998wa} N.~Kaiser, S.~Gerstendorfer, and W.~Weise, 
%``Peripheral N N scattering: Role of Delta excitation, correlated  two-pion and vector meson exchange,''
Nucl.\ Phys.\ \textbf{A637}, 395 (1998).% [arXiv:nucl-th/9802071]. 
%%CITATION = NUPHA,A637,395;%%

\bibitem{Krebs:2007rh}
  H.~Krebs, E.~Epelbaum, and U.~-G.~Meissner,
  %``Nuclear forces with Delta-excitations up to next-to-next-to-leading order. I. Peripheral nucleon-nucleon waves,''
  Eur.\ Phys.\ J.\  {\bf A32}, 127 (2007).

\bibitem{Pascalutsa:2002pi}
  V.~Pascalutsa and D.~R.~Phillips,
  %``Effective theory of the delta(1232) in Compton scattering off the nucleon,''
  Phys.\ Rev.\  {\bf C67}, 055202 (2003).
%   [nucl-th/0212024].

\bibitem{Long:2009wq}
  Bingwei~Long and U.~van Kolck,
  %``pi N Scattering in the Delta(1232) Region in an Effective Field Theory,''
  Nucl.\ Phys.\  {\bf A840}, 39 (2010).
%   [arXiv:0907.4569 [hep-ph]].
% Nucl.\ Phys.\ {\bf A}, in press, [arXiv:1105.2764 [nucl-th]].

\bibitem{Long:2010kt}
  Bingwei~Long and V.~Lensky,
  %``Heavy-particle formalism with Foldy-Wouthuysen representation,''
  Phys.\ Rev.\  {\bf C83}, 045206 (2011).
%   [arXiv:1010.2738 [hep-ph]].


\bibitem{Valderrama:2008kj_2010fb} 
  M.~P.~Valderrama and E.~Ruiz Arriola,
  %``Renormalization of chiral two-pion exchange NN interactions with Delta-excitations: Central Phases and the Deuteron,''
  Phys.\ Rev.\ {\bf C79}, 044001 (2009);
%   [arXiv:0809.3186 [nucl-th]].
  %%CITATION = ARXIV:0809.3186;%%
% \bibitem{PavonValderrama:2010fb} 
%   M.~Pavon Valderrama and E.~Ruiz Arriola,
  %``Renormalization of chiral two pion exchange NN interactions with Delta-excitations: Correlations in the partial wave expansion,''
  \textit{ibid}. {\bf C83}, 044002 (2011).
%   [arXiv:1005.0744 [nucl-th]].
  %%CITATION = ARXIV:1005.0744;%%

\bibitem{Entem:2007jg} D.~R.~Entem, E.~Ruiz Arriola, M.~Pavon Valderrama and
R.~Machleidt, 
%``Renormalization of chiral two-pion exchange NN interactions. momentum versus coordinate space,''
Phys.\ Rev.\ \textbf{C77}, 044006 (2008). %[arXiv:0709.2770 [nucl-th]].


\bibitem{Pandharipande:2005sx} 
  V.~R.~Pandharipande, D.~R.~Phillips, and U.~van Kolck,
  %``Delta effects in pion-nucleon scattering and the strength of the two-pion-exchange three-nucleon interaction,''
  Phys.\ Rev.\ {\bf C71}, 064002 (2005).
%   [nucl-th/0501061].
  %%CITATION = NUCL-TH/0501061;%%


\bibitem{Baru:2012iv} 
  V.~Baru, E.~Epelbaum, C.~Hanhart, M.~Hoferichter, A.~E.~Kudryavtsev, and D.~R.~Phillips,
  %``The Multiple-scattering series in pion-deuteron scattering and the nucleon-nucleon potential: Perspectives from effective field theory,''
  Eur. Phys. J. A \textbf{48}, 69 (2012).
  %%CITATION = ARXIV:1202.0208;%%

\bibitem{bira-pionless}
U. van Kolck,
in \textit{Proceedings of the Workshop on Chiral Dynamics 1997, Theory
  and Experiment,} edited by A. Bernstein, D. Drechsel, and T. Walcher
(Spring-Verlag, Berlin, 1998);
Nucl.\ Phys.\ {\bf A645}, 273 (1999).

\bibitem{Gegelia} 
  J.~Gegelia,
  %``Nucleon-nucleon scattering and effective field theory: Including pions nonperturbatively,''
  Phys.\ Lett.\ {\bf B463}, 133 (1999);
%   [nucl-th/9908055].
  J.~Gegelia and G.~Japaridze,
  %``Renormalization of singlet S wave N N scattering amplitude in effective field theory,''
  ibid. {\bf B517}, 476 (2001).


\end{thebibliography}
\end{document}